\documentclass[twocolumn]{aastex61}
\usepackage{amsmath,amstext}

\def \ucla {Department of Physics and Astronomy, University of California, Los Angeles, 475 Portola Plaza, Los Angeles, CA 90095-1547, USA}
\def \ethOld {Department of Physics, ETH Zurich, Wolfgang-Pauli-Strasse 27, 8093, Zurich, Switzerland}
\def \eth {Institute of Particle Physics and Astrophysics, Department of Physics, ETH Zurich, Wolfgang-Pauli-Strasse 27, 8093, Zurich, Switzerland}

\submitjournal{ApJL}

\shorttitle{Cosmic Shear with Einstein Rings}
\shortauthors{Birrer, Refregier \& Amara}

\begin{document}

\title{Cosmic Shear with Einstein Rings}

\correspondingauthor{Simon Birrer}
\email{sibirrer@astro.ucla.edu, alexandre.refregier@phys.ethz.ch, adam.amara@phys.ethz.ch}
\author[0000-0003-3195-5507]{Simon Birrer}
\affiliation{\ucla}
\affiliation{\ethOld}
\author{Alexandre Refregier}
\affiliation{\eth}
\author{Adam Amara}
\affiliation{\eth}

\begin{abstract}

We explore a new technique to measure cosmic shear using Einstein rings. In \cite{Birrer:2017los}, we showed that the detailed modelling of Einstein rings can be used to measure external shear to high precision. In this letter, we explore how a collection of Einstein rings can be used as a statistical probe of cosmic shear. We present a forecast of the cosmic shear information available in Einstein rings for different strong lensing survey configurations. We find that, assuming that the number density of Einstein rings in the COSMOS survey is representative, future strong lensing surveys should have a cosmological precision comparable to the current ground based weak lensing surveys. We discuss how this technique is complementary to the standard cosmic shear analyses since it is sensitive to different systematic and can be used for cross-calibration.

\end{abstract}


\section{Introduction}
Probes of cosmic large-scale structures of the universe are powerful tools for testing dark matter, dark energy and gravity on large scales \citep[see e.g.][]{Weinberg:2013a}. One such probe is cosmic shear, which can be used to measure density perturbations through the distortions of galaxy images \citep[][and references therein]{Kaiser:1992, Bartelmann:2001vu, Refregier:2003xq, Albrecht:2006a, Peacock:2006a, Hoekstra:2008a, Munshi:2008a, Massey:2010a}. Since the first detections of cosmic shear in 2000 \citep{Bacon:2000a, Kaiser:2000a, Wittman:2000a, VanWaerbeke:2000}, several surveys have measured the correlation of galaxy shapes on cosmological scales, such as the Sloan Digital Sky Survey \citep{Lin:2012a, Huff:2014a}, the Deep Lens Survey \citep{Jee:2013a}, the Canada-France-Hawaii Legacy Survey \citep{Kilbinger:2013a, Heymans:2013a}, the Kilo Degree Survey \citep{Kuijken:2015a} and the Dark Energy Survey \citep{Becker:2016a, DES_shear_Y1}.

The intrinsic galaxy ellipticities are about one order of magnitude larger than those induced by cosmic shear. As a result, one needs to average over a large ensemble of galaxies to measure the sought-after weak lensing signal. The low signal-to-noise ratio per galaxy also makes the measurements sensitive to systematic effects such as uncertainties in the point spread function, the noise properties, etc. With the increase in the depth and area of cosmic shear surveys, control of systematics continues to be a major focus in their cosmological analyses \citep[see e.g.][]{Amara:2008}.

Galaxy shapes are not the only way to measure large scale structure through gravitational lensing. Cosmic magnification \citep{Menard:2002} has been proposed and statistically detected in SDSS \citep{Scranton:2005}. Lensing magnification of supernovae of type Ia may also provide an independent measurement. The intrinsic scatter in SNIa is about 10-15\% \citep[][]{Cooray:2006a}, which is higher than the magnification signal imprinted. Weak lensing of the Lyman-alpha forest \citep[][]{Croft:2017} has also been proposed to measure the weak lensing effect.

In the modeling of strong lens systems, in particular quadruply lensed quasars, weak lensing distortions have to be modeled to match the observables. This has been done by introducing linear shear terms in additon to the main deflector model \citep[see e.g.][]{Hogg:1994, Keeton:1997, Schechter:1997, Fischer:1998, Kochanek:2001, Suyu:2013ni, Birrer:2016zy} or by explicitly modeling the nearby dominant galaxies and their dark matter halo \citep[][]{Wong:2017a}.

Recently, we have demonstrated in \cite{Birrer:2017los} that the careful forward modelling and introduction of non-linear shear terms acting on the main deflector of strong lens systems can provide a precise measurement of the external weak lensing shear. In particular, the analysis of a single Einstein ring system in the COSMOS field yielded a shear precision of $\pm 0.003$ for both, the shear acting on the main deflector and the integrated shear to the souce plane.

In this letter, we explore how a collection of Einstein rings can be used as a statistical probe of cosmic shear. We first review how Einstein rings can be used for this purpose (\S\ref{sec:shear_measurement}). We then forecast the statistical sensitivity of Einstein ring surveys for cosmic shear and compare it to that of galaxy shape surveys (\S\ref{sec:forecast}). After discussing the complementarity of these two approaches (\S\ref{sec:discussion}) we summarise our conclusions (\S\ref{sec:conclusion}).

\section{Shear measurements from Einstein rings}\label{sec:shear_measurement}

In \cite{Birrer:2017los}, we presented a method to accurately model the effects of large scale structures along the line of sight (LOS) on strong lensing systems. This forward modeling approach allowed us to separate the weak lensing LOS shear effect from the main strong lens deflector.

The LOS effect can generally be modelled with four distortion parameters: Two reduced shear terms that describe the weak lensing effect between the main deflector and the observer and two shear terms that describe the integrated shear terms between the source and the observer via a non-linear path through the main deflector. These effects are mathematically distinct from ellipticity in the main deflector.

An idealised case is an Einstein ring with perfect circular lens and a point source. In this case the presence of large scale structure along the LOS alters the shape of the Einstein ring. In particular the ring becomes elliptical due to weak lensing by foreground structures. Since an elliptical Einstein ring can not be produced by an elliptical lens model, it can be considered as a signature of cosmic shear along the LOS and the Einstein ring as a \textit{standard shape}.

\begin{figure*}
  \centering
  \includegraphics[angle=0, width=160mm]{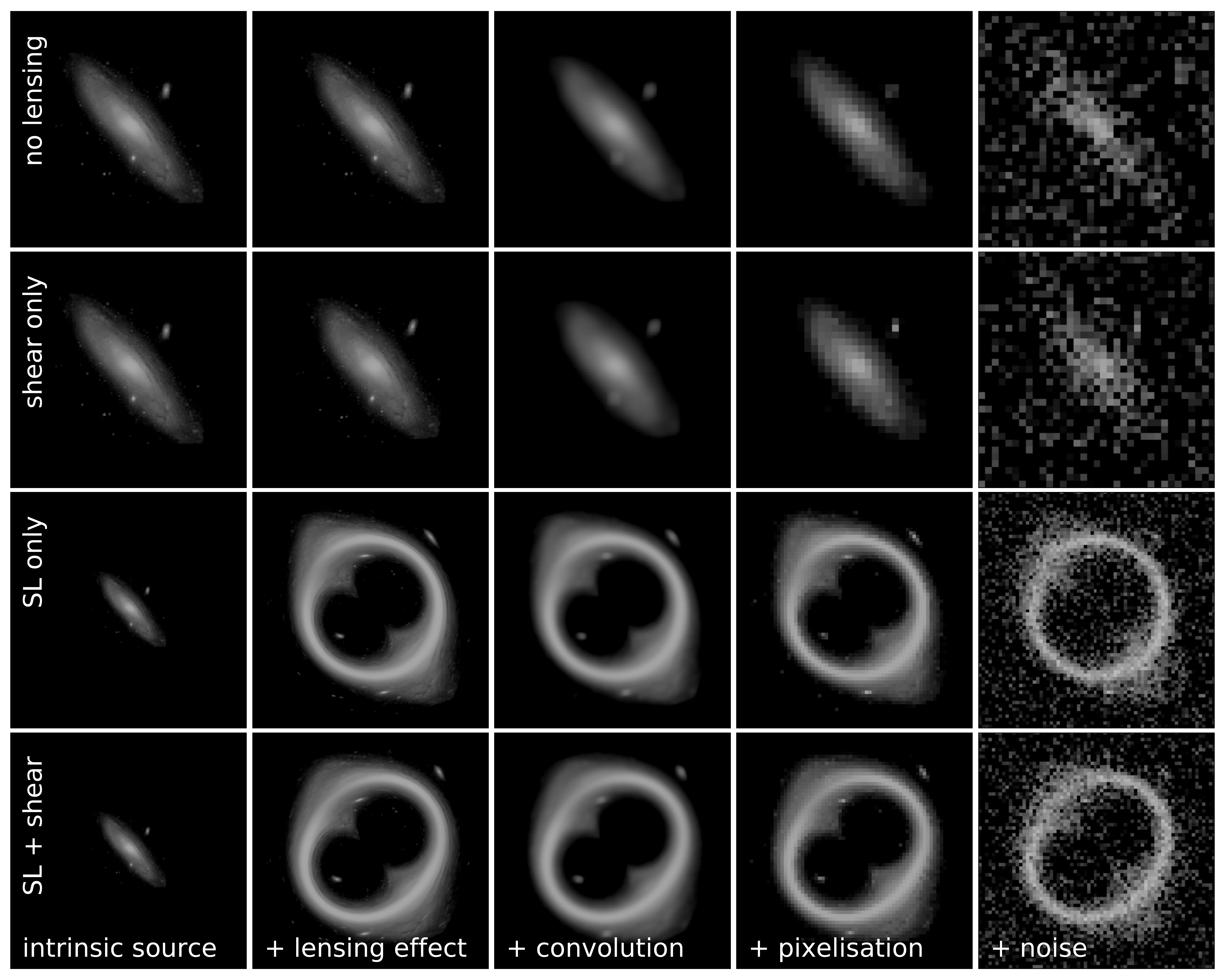}
  \caption{Illustration of Einstein rings to measure cosmic shear. \textbf{From left to right:} Forward modeling of a distant galaxy (1) through gravitational lensing (2), diffraction and atmospheric effects leading to a convolution (FWHM 0.1") (3), detector discreteness (pixelised grid of 0.05"$\times$0.05") and noise (5). \textbf{From top to bottom:} No lensing (1), only cosmic shear with amplitude 0.1 (2), a strong lens with Einstein radius $\theta_{\text{E}}=1"$ (3) and a strong lens with an additional cosmic shear in the foreground of the lens corresponding to a shear at the lens plane of 0.06 and at the source plane of 0.1 (4). In the standard galaxy weak lensing (second row), the shear term can only be inferred when the intrinsic source is known and observational conditions degrade this effect with potential biases (comparison of first and second raw). In the strong lensing regime, the direction and strength of the shear has a reliably measurable imprint in the shape of an Einstein ring (bottom right).}
\label{fig:illustration}
\end{figure*}

Figure \ref{fig:illustration} demonstrates the use of Einstein rings for measuring cosmic shear. The forward modelling of a distant galaxy is illustrated through 4 different cases: (1) without lensing, (2) with cosmic shear only, (3) with a strong lens only and (4) with both cosmic shear and strong lensing. In the standard galaxy weak lensing, the effect of shear is subdominant compared to the intrinsic shape of galaxies (see rows 1 and 2 of the figure). In addition, observational effects such as convolution by the PSF, pixelization and noise further degrade the available shear information per galaxy (see columns 3 to 5). In the strong lensing regime, the direction and amplitude of the shear has a noticeable imprint in the distortion of an Einstein ring (see rows 3 and 4).

In practice, the cosmic shear information imprinted in strong lens imaging data can be extracted by forward modelling and simultaneously reconstructing the strong lens (including cosmic shear terms) and the source surface brightness \citep[see ][for further details]{Birrer:2017los}. 

We tested this approach in \cite{Birrer:2017los} using mock data and we showed that strong lens systems can accurately measure cosmic shear with a precision of $\sigma_{\gamma, \text{SL}} = \pm 0.003$ (statistical error) for an HST-like dataset of an Einstein ring. We also achieved a similar precision in the reconstruction of the Einstein ring lens COSMOS 0038+4133. The uncertainties quoted include the marginalization over lens ellipticity parameters, source position and surface brightness distribution. For this letter, we provide further test cases for a broader range of Einstein rings in the form of a jupyter notebook
\footnote{\url{http://www.astro.ucla.edu/~sibirrer/CosmicShearEinsteinRing/}} and we also provide the open-source software \textit{lenstronomy} (Birrer et al. in prep) \footnote{\url{https://github.com/sibirrer/lenstronomy}}. In particular, we provide tests with moderate ellipticity in the main deflector that leads to a split in the image configuration. For those tests we achieve the same cosmic shear precision of $\sigma_{\gamma, \text{SL}} = \pm 0.003$.

\section{Cosmic shear forecast}\label{sec:forecast}

For weak lensing galaxy surveys, we estimate the shear uncertainty variance for a unit area by \citep[see e.g.][]{Amara:2008}, 

\begin{equation} \label{eqn:sigma_hat_WL}
	\hat{\sigma}^2_{\gamma, \text{gal}} = \frac{\sigma_{\gamma, \text{gal}}^2}{n_{\text{gal}}},
\end{equation}
where  $\sigma_{\gamma, \text{gal}}^2$ is the shear uncertainty variance per galaxy and $n_{\text{gal}}$ is the galaxy surface number density. Similarly, the shear uncertainty variance for an Einstein ring survey is given by
\begin{equation}\label{eqn:sigma_hat_SL}
	\hat{\sigma}_{\gamma, \text{SL}}^2 = \frac{\sigma_{\gamma, \text{SL}}^2}{n_{\text{SL}}},
\end{equation}
where $\sigma_{\gamma, \text{SL}}$ is the shear uncertainty per Einstein ring and $n_{\text{SL}}$ is the number density of Einstein rings. Note that all shear uncertainties are given per shear component.

To estimate the number density $n_{\text{SL}}$ of Einstein rings that can be modeled with sufficient precision, we consider the strong lens systems found in the COSMOS survey \cite{Scoville:2007a}.
The COSMOS survey is based on HST imaging over an area of 1.64 deg$^2$ with a limiting magnitude of about $I_{\text{F814W}} = 26.5$ mag (10-$\sigma$ point source detection). The systematic search for strong gravitational lenses \citep{Faure:2008a} resulted in 20 strong lens systems with multiple images or large curved arcs. For a few of their strong lenses, the authors were able to infer the external shear with forward modeling. For the present study, we focus on the few systems where reliable and high precision cosmic shear values can be extracted. In our previous study \citep[][]{Birrer:2017los}, we used one Einstein ring lens system, COSMOS 0038+4133, to measure the external shear to high precision. There is at least one other valuable strong lens in the COSMOS field, that is nearly an Einstein ring. We thus take $n_{\text{SL}} \approx 1$ deg$^{-2}$ for the forecast presented in this letter.

A useful comparison of the cosmic shear information is given by the dark energy Figure of Merit (FoM) \citep[][]{Albrecht:2006a}. We follow the analysis of \cite{Amara:2007a} and take their equation (10) for the scaling of the FoM as a function of weak lensing survey parameters. To adapt this scaling to also apply to Einstein rings, we include an explicit scaling for the shear precision $\sigma_{\gamma,\text{gal/SL}}$. For simplicity we do not consider other scaling factors.  This results in a dark energy FoM forecast that depends on survey area $A_{\text{s}}$, number density of sources $n_{\text{gal/SL}}$ and shear precision per source $\sigma_{\gamma, \text{gal/SL}}$
\begin{equation} \label{eqn:fom}
	\text{FoM}_{\text{DE}} \approx 2.8 \left(\frac{A_{\text{s}}}{5\times 10^3 {\rm ~deg}^2} \right) \left(\frac{n_{\text{gal/SL}}}{10 ~{\rm arcmin}^{-2}} \right) \left(\frac{\sigma_{\gamma ,\text{gal/SL}}}{0.25} \right)^{-2}.
\end{equation}

For the strong lensing forecast, we estimate the abundance and precision from the COSMOS lens sample with $n_{\text{SL}} \approx 1$ deg$^{-2}$ and $\sigma_{\gamma, \text{SL}} = 0.003$, as discussed above. We compute the FoM for two different areas, survey SL1 with 5000 deg$^2$ and SL2 with 20000 deg$^2$. We compare the strong lensing information to two different weak lensing survey configurations. For both surveys we assume a combined shape noise and measurement noise of $\sigma_{\gamma, \text{gal}} = 0.25$. For the first survey WL1, we choose a set of parameters that is comparable to current weak lensing measurements, namely  $A_{s} = 1500$ deg$^2$ and $n_{\text{gal}} = 5$ per arcmin$^{-2}$. For the second one, WL2, we choose $n_{\text{gal}} = 10$ arcmin$^{-2}$ and the same area of 5000 deg$^2$ as SL1. 

Table \ref{tab:forecast_table} summarizes the forecasts for the different surveys and provides estimates of the noise density of shear measurement for a unit area, and for the FoM of the dark energy equation of state (Equation \ref{eqn:fom}). For WL2, we get shear uncertainties $\hat{\sigma}_{\gamma, \text{WL}}^2 \approx 2.5 \times 10^{-6}$ deg$^{2}$. The strong lenses are estimated to provide $\hat{\sigma}_{\gamma, \text{SL}}^2 \approx 9 \times 10^{-6}$ deg$^{2}$, a number comparable to the weak lensing. The figure of merit shows that a strong lens survey of 5000 deg$^2$ has a similar performance as existing weak lensing surveys and that when extending the area to 20000 deg$^2$, strong lensing will be comparable with the 5000 deg$^2$ survey WL2.

\begin{deluxetable*}{ l | c | c | c | c | c | c } 
\tablecaption{Cosmic shear forecast for weak lensing and strong lensing surveys\label{tab:forecast_table}}
\tablewidth{0pt}
\tablehead{
Survey & $A_{s}$ [deg$^2$]& $n_\text{gal/SL}$ & $\sigma_{\gamma,\text{gal/SL}}$ & $N_{tot}$ & $\hat{\sigma}_{\gamma,\text{gal/SL}}^2$  [deg$^{2}$] & FoM$_{DE}$ 
}
\startdata
SL1&  5000  & 1 deg$^{-2}$ & 0.003 & 5.0e+03 & 9.0e-06 & 0.50 \\
SL2&  20000  & 1 deg$^{-2}$ & 0.003 & 2.0e+04 & 9.0e-06 & 2.2 \\
WL1&  1500  & 5 arcmin$^{-2}$ & 0.25 & 2.4e+07 & 3.5e-06 & 0.42 \\
WL2&  5000  & 10 arcmin$^{-2}$ & 0.25 & 1.8e+08 & 1.7e-06 & 2.8 \\
\enddata
\tablecomments{Cosmic shear forecast for weak lensing and strong lensing surveys of different configurations in terms of area $A_{s}$, number density of galaxies/strong lenses $n_\text{gal/SL}$ and shear error per object $\sigma_{\gamma,\text{gal/SL}}$. Weak lensing surveys have about $10^4$ times more sources, $N_{tot}$, than strong lens surveys but their shear variance for a unit area, $\hat{\sigma}_{\gamma,\text{gal/SL}}^2$, is only a factor 2-3 superior. The figure of merit of the dark energy equation of state FoM$_{DE}$ is computed according to Equation \ref{eqn:fom}.}

\end{deluxetable*}

\section{Discussion}\label{sec:discussion}

The cosmic shear forecast for the different surveys in Section \ref{sec:forecast} do not include systematic limitations, neither for the galaxy shape surveys nor for the Einstein ring forecast. \cite{Amara:2007a} finds a steep decrease in the FoM of dark energy for weak lensing surveys when increasing the uncertainty in the redshift estimate or the fraction of catastrophic failures. Current and future galaxy surveys rely on photometric redshifts since the number and depth of the galaxy sample prohibits a full complete spectroscopic follow up. On the other hand, the number of Einstein rings for cosmic shear studies is within the range of a full spectroscopic follow up, thereby alleviating the limitations from photometric redshift estimation.

The cosmic shear estimate from Einstein rings may however also be affected by systematics. For general strong lens systems, the strong lens model can be partially degenerate with cosmic shear. In the present analysis and in \cite{Birrer:2017los}, we focus on Einstein ring systems. The simpler and more circular the lens, the more directly the shape of an Einstein ring can be attributed to cosmic shear and the inference of the shear parameters become more precise. A detailed study of the distribution of realistic lenses and their infered precision on the cosmic shear parameters is beyond the scope of this letter.

The strong lensing systems are also likely to have selection biases favouring high density regions, an affect that would need to be modelled \citep[e.g.][]{Holder:2003}. Detailed checks of potential systematics may be performed for a specific set of lenses \citep[see e.g.][for the impact of the source scale and mass-sheet degeneracy]{Birrer:2016zy}.

Galaxy shapes and Einstein ring measurements of cosmic shear are complementary. They are effected by different systematics but probe the same LOS density field. Einstein rings can make precise measurements of cosmic shear at a few locations, while galaxy shapes provide a large number of measurements homogeneously over the survey, but at much lower signal-to-noise. A joint analysis of Einstein rings and galaxy shapes can allow for cross-correlations and cross-calibrations.

\section{Conclusion}\label{sec:conclusion}
In this letter, we have proposed Einstein rings as ``standard shapes" to measure cosmic shear. We presented forecasts for the cosmic shear sensitivity of future Einstein ring surveys. We found that their sensitivity is comparable to that of current weak lensing surveys, assuming that the number density of Einstein rings in the COSMOS survey is representative. Einstein rings probe the same large scale structure as standard cosmic shear analyses, but are effected by different systematics. They are therefore an independent probe of cosmic shear that is complementary to galaxy shape surveys.

\acknowledgments
SB thanks Tommaso Treu for support and funding. We thank the anonymeous referee for useful comments that clarified the presentation of this work.

\bibliographystyle{apj}

\begin{thebibliography}{}
\expandafter\ifx\csname natexlab\endcsname\relax\def\natexlab#1{#1}\fi

\bibitem[{{Albrecht} {et~al.}(2006){Albrecht}, {Bernstein}, {Cahn}, {Freedman},
  {Hewitt}, {Hu}, {Huth}, {Kamionkowski}, {Kolb}, {Knox}, {Mather}, {Staggs},
  \& {Suntzeff}}]{Albrecht:2006a}
{Albrecht}, A., {Bernstein}, G., {Cahn}, R., {et~al.} 2006, ArXiv Astrophysics
  e-prints, astro-ph/0609591

\bibitem[{{Amara} \& {R{\'e}fr{\'e}gier}(2007)}]{Amara:2007a}
{Amara}, A., \& {R{\'e}fr{\'e}gier}, A. 2007, \mnras, 381, 1018

\bibitem[{{Amara} \& {R{\'e}fr{\'e}gier}(2008)}]{Amara:2008}
---. 2008, \mnras, 391, 228

\bibitem[{{Bacon} {et~al.}(2000){Bacon}, {Refregier}, \& {Ellis}}]{Bacon:2000a}
{Bacon}, D.~J., {Refregier}, A.~R., \& {Ellis}, R.~S. 2000, \mnras, 318, 625

\bibitem[{Bartelmann \& Schneider(2001)}]{Bartelmann:2001vu}
Bartelmann, M., \& Schneider, P. 2001, Physics Reports, 340, 291

\bibitem[{{Becker} {et~al.}(2016){Becker}, {Troxel}, {MacCrann}, {Krause},
  {Eifler}, {Friedrich}, {Nicola}, {Refregier}, {Amara}, {Bacon}, {Bernstein},
  {Bonnett}, {Bridle}, {Busha}, {Chang}, {Dodelson}, {Erickson}, {Evrard},
  {Frieman}, {Gaztanaga}, {Gruen}, {Hartley}, {Jain}, {Jarvis}, {Kacprzak},
  {Kirk}, {Kravtsov}, {Leistedt}, {Peiris}, {Rykoff}, {Sabiu}, {S{\'a}nchez},
  {Seo}, {Sheldon}, {Wechsler}, {Zuntz}, {Abbott}, {Abdalla}, {Allam},
  {Armstrong}, {Banerji}, {Bauer}, {Benoit-L{\'e}vy}, {Bertin}, {Brooks},
  {Buckley-Geer}, {Burke}, {Capozzi}, {Carnero Rosell}, {Carrasco Kind},
  {Carretero}, {Castander}, {Crocce}, {Cunha}, {D'Andrea}, {da Costa}, {DePoy},
  {Desai}, {Diehl}, {Dietrich}, {Doel}, {Fausti Neto}, {Fernandez}, {Finley},
  {Flaugher}, {Fosalba}, {Gerdes}, {Gruendl}, {Gutierrez}, {Honscheid},
  {James}, {Kuehn}, {Kuropatkin}, {Lahav}, {Li}, {Lima}, {Maia}, {March},
  {Martini}, {Melchior}, {Miller}, {Miquel}, {Mohr}, {Nichol}, {Nord},
  {Ogando}, {Plazas}, {Reil}, {Romer}, {Roodman}, {Sako}, {Sanchez},
  {Scarpine}, {Schubnell}, {Sevilla-Noarbe}, {Smith}, {Soares-Santos},
  {Sobreira}, {Suchyta}, {Swanson}, {Tarle}, {Thaler}, {Thomas}, {Vikram},
  {Walker}, \& {Dark Energy Survey Collaboration}}]{Becker:2016a}
{Becker}, M.~R., {Troxel}, M.~A., {MacCrann}, N., {et~al.} 2016, \prd, 94,
  022002

\bibitem[{Birrer {et~al.}(2016)Birrer, Amara, \& Refregier}]{Birrer:2016zy}
Birrer, S., Amara, A., \& Refregier, A. 2016, Journal of Cosmology and
  Astroparticle Physics, 08, 020

\bibitem[{{Birrer} {et~al.}(2017){Birrer}, {Welschen}, {Amara}, \&
  {Refregier}}]{Birrer:2017los}
{Birrer}, S., {Welschen}, C., {Amara}, A., \& {Refregier}, A. 2017, \jcap, 4,
  049

\bibitem[{{Cooray} {et~al.}(2006){Cooray}, {Holz}, \& {Huterer}}]{Cooray:2006a}
{Cooray}, A., {Holz}, D.~E., \& {Huterer}, D. 2006, \apjl, 637, L77

\bibitem[{{Croft} {et~al.}(2017){Croft}, {Romeo}, \& {Metcalf}}]{Croft:2017}
{Croft}, R.~A.~C., {Romeo}, A., \& {Metcalf}, R.~B. 2017, ArXiv e-prints,
  arXiv:1706.07870

\bibitem[{{Faure} {et~al.}(2008){Faure}, {Kneib}, {Covone}, {Tasca},
  {Leauthaud}, {Capak}, {Jahnke}, {Smolcic}, {de la Torre}, {Ellis},
  {Finoguenov}, {Koekemoer}, {Le Fevre}, {Massey}, {Mellier}, {Refregier},
  {Rhodes}, {Scoville}, {Schinnerer}, {Taylor}, {Van Waerbeke}, \&
  {Walcher}}]{Faure:2008a}
{Faure}, C., {Kneib}, J.-P., {Covone}, G., {et~al.} 2008, \apjs, 176, 19

\bibitem[{{Fischer} {et~al.}(1998){Fischer}, {Schade}, \&
  {Barrientos}}]{Fischer:1998}
{Fischer}, P., {Schade}, D., \& {Barrientos}, L.~F. 1998, \apjl, 503, L127

\bibitem[{{Heymans} {et~al.}(2013){Heymans}, {Grocutt}, {Heavens}, {Kilbinger},
  {Kitching}, {Simpson}, {Benjamin}, {Erben}, {Hildebrandt}, {Hoekstra},
  {Mellier}, {Miller}, {Van Waerbeke}, {Brown}, {Coupon}, {Fu},
  {Harnois-D{\'e}raps}, {Hudson}, {Kuijken}, {Rowe}, {Schrabback}, {Semboloni},
  {Vafaei}, \& {Velander}}]{Heymans:2013a}
{Heymans}, C., {Grocutt}, E., {Heavens}, A., {et~al.} 2013, \mnras, 432, 2433

\bibitem[{{Hoekstra} \& {Jain}(2008)}]{Hoekstra:2008a}
{Hoekstra}, H., \& {Jain}, B. 2008, Annual Review of Nuclear and Particle
  Science, 58, 99

\bibitem[{{Hogg} \& {Blandford}(1994)}]{Hogg:1994}
{Hogg}, D.~W., \& {Blandford}, R.~D. 1994, \mnras, 268, 889

\bibitem[{{Holder} \& {Schechter}(2003)}]{Holder:2003}
{Holder}, G.~P., \& {Schechter}, P.~L. 2003, \apj, 589, 688

\bibitem[{{Huff} {et~al.}(2014){Huff}, {Eifler}, {Hirata}, {Mandelbaum},
  {Schlegel}, \& {Seljak}}]{Huff:2014a}
{Huff}, E.~M., {Eifler}, T., {Hirata}, C.~M., {et~al.} 2014, \mnras, 440, 1322

\bibitem[{{Jee} {et~al.}(2013){Jee}, {Tyson}, {Schneider}, {Wittman},
  {Schmidt}, \& {Hilbert}}]{Jee:2013a}
{Jee}, M.~J., {Tyson}, J.~A., {Schneider}, M.~D., {et~al.} 2013, \apj, 765, 74

\bibitem[{{Kaiser}(1992)}]{Kaiser:1992}
{Kaiser}, N. 1992, \apj, 388, 272

\bibitem[{{Kaiser} {et~al.}(2000){Kaiser}, {Wilson}, \&
  {Luppino}}]{Kaiser:2000a}
{Kaiser}, N., {Wilson}, G., \& {Luppino}, G.~A. 2000, ArXiv Astrophysics
  e-prints, astro-ph/0003338

\bibitem[{{Keeton} {et~al.}(1997){Keeton}, {Kochanek}, \&
  {Seljak}}]{Keeton:1997}
{Keeton}, C.~R., {Kochanek}, C.~S., \& {Seljak}, U. 1997, \apj, 482, 604

\bibitem[{{Kilbinger} {et~al.}(2013){Kilbinger}, {Fu}, {Heymans}, {Simpson},
  {Benjamin}, {Erben}, {Harnois-D{\'e}raps}, {Hoekstra}, {Hildebrandt},
  {Kitching}, {Mellier}, {Miller}, {Van Waerbeke}, {Benabed}, {Bonnett},
  {Coupon}, {Hudson}, {Kuijken}, {Rowe}, {Schrabback}, {Semboloni}, {Vafaei},
  \& {Velander}}]{Kilbinger:2013a}
{Kilbinger}, M., {Fu}, L., {Heymans}, C., {et~al.} 2013, \mnras, 430, 2200

\bibitem[{{Kochanek} {et~al.}(2001){Kochanek}, {Keeton}, \&
  {McLeod}}]{Kochanek:2001}
{Kochanek}, C.~S. and {Keeton}, C.~R. and {McLeod}, B.~A. 2001, \apj, 547, 50

\bibitem[{{Kuijken} {et~al.}(2015){Kuijken}, {Heymans}, {Hildebrandt},
  {Nakajima}, {Erben}, {de Jong}, {Viola}, {Choi}, {Hoekstra}, {Miller}, {van
  Uitert}, {Amon}, {Blake}, {Brouwer}, {Buddendiek}, {Conti}, {Eriksen},
  {Grado}, {Harnois-D{\'e}raps}, {Helmich}, {Herbonnet}, {Irisarri},
  {Kitching}, {Klaes}, {La Barbera}, {Napolitano}, {Radovich}, {Schneider},
  {Sif{\'o}n}, {Sikkema}, {Simon}, {Tudorica}, {Valentijn}, {Verdoes Kleijn},
  \& {van Waerbeke}}]{Kuijken:2015a}
{Kuijken}, K., {Heymans}, C., {Hildebrandt}, H., {et~al.} 2015, \mnras, 454,
  3500

\bibitem[{{Lin} {et~al.}(2012){Lin}, {Dodelson}, {Seo}, {Soares-Santos},
  {Annis}, {Hao}, {Johnston}, {Kubo}, {Reis}, \& {Simet}}]{Lin:2012a}
{Lin}, H., {Dodelson}, S., {Seo}, H.-J., {et~al.} 2012, \apj, 761, 15

\bibitem[{{Massey} {et~al.}(2010){Massey}, {Kitching}, \&
  {Richard}}]{Massey:2010a}
{Massey}, R., {Kitching}, T., \& {Richard}, J. 2010, Reports on Progress in
  Physics, 73, 086901

\bibitem[{{M{\'e}nard} \& {Bartelmann}(2002)}]{Menard:2002}
{M{\'e}nard}, B., \& {Bartelmann}, M. 2002, \aap, 386, 784

\bibitem[{{Munshi} {et~al.}(2008){Munshi}, {Valageas}, {van Waerbeke}, \&
  {Heavens}}]{Munshi:2008a}
{Munshi}, D., {Valageas}, P., {van Waerbeke}, L., \& {Heavens}, A. 2008,
  \physrep, 462, 67

\bibitem[{{Peacock} {et~al.}(2006){Peacock}, {Schneider}, {Efstathiou},
  {Ellis}, {Leibundgut}, {Lilly}, \& {Mellier}}]{Peacock:2006a}
{Peacock}, J.~A., {Schneider}, P., {Efstathiou}, G., {et~al.} 2006, {ESA-ESO
  Working Group on ``Fundamental Cosmology''}, Tech. rep., astro-ph/0610906

\bibitem[{Refregier(2003)}]{Refregier:2003xq}
Refregier, A. 2003, Annual Review of Astronomy \&Astrophysics, 41, 645

\bibitem[{{Schechter} {et~al.}(1997){Schechter}, {Bailyn}, {Barr}, {Barvainis},
  {Becker}, {Bernstein}, {Blakeslee}, {Bus}, {Dressler}, {Falco}, {Fesen},
  {Fischer}, {Gebhardt}, {Harmer}, {Hewitt}, {Hjorth}, {Hurt}, {Jaunsen},
  {Mateo}, {Mehlert}, {Richstone}, {Sparke}, {Thorstensen}, {Tonry}, {Wegner},
  {Willmarth}, \& {Worthey}}]{Schechter:1997}
{Schechter}, P.~L., {Bailyn}, C.~D., {Barr}, R., {et~al.} 1997, \apjl, 475, L85

\bibitem[{{Scoville} {et~al.}(2007){Scoville}, {Abraham}, {Aussel}, {Barnes},
  {Benson}, {Blain}, {Calzetti}, {Comastri}, {Capak}, {Carilli}, {Carlstrom},
  {Carollo}, {Colbert}, {Daddi}, {Ellis}, {Elvis}, {Ewald}, {Fall},
  {Franceschini}, {Giavalisco}, {Green}, {Griffiths}, {Guzzo}, {Hasinger},
  {Impey}, {Kneib}, {Koda}, {Koekemoer}, {Lefevre}, {Lilly}, {Liu},
  {McCracken}, {Massey}, {Mellier}, {Miyazaki}, {Mobasher}, {Mould}, {Norman},
  {Refregier}, {Renzini}, {Rhodes}, {Rich}, {Sanders}, {Schiminovich},
  {Schinnerer}, {Scodeggio}, {Sheth}, {Shopbell}, {Taniguchi}, {Tyson}, {Urry},
  {Van Waerbeke}, {Vettolani}, {White}, \& {Yan}}]{Scoville:2007a}
{Scoville}, N., {Abraham}, R.~G., {Aussel}, H., {et~al.} 2007, \apjs, 172, 38

\bibitem[{{Scranton} {et~al.}(2005){Scranton}, {M{\'e}nard}, {Richards},
  {Nichol}, {Myers}, {Jain}, {Gray}, {Bartelmann}, {Brunner}, {Connolly},
  {Gunn}, {Sheth}, {Bahcall}, {Brinkman}, {Loveday}, {Schneider}, {Thakar}, \&
  {York}}]{Scranton:2005}
{Scranton}, R., {M{\'e}nard}, B., {Richards}, G.~T., {et~al.} 2005, \apj, 633,
  589

\bibitem[{Suyu {et~al.}(2013)Suyu, Auger, Hilbert, Marshall, Tewes, Treu,
  Fassnacht, Koopmans, Sluse, Blandford, Courbin, \& Meylan}]{Suyu:2013ni}
Suyu, S., Auger, M., Hilbert, S., {et~al.} 2013, The Astrophysical Journal,
  766, 70

\bibitem[{{Troxel} {et~al.}(2017){Troxel}, {MacCrann}, {Zuntz}, {Eifler},
  {Krause}, {Dodelson}, {Gruen}, {Blazek}, {Friedrich}, {Samuroff}, {Prat},
  {Secco}, {Davis}, {Fert{\'e}}, {DeRose}, {Alarcon}, {Amara}, {Baxter},
  {Becker}, {Bernstein}, {Bridle}, {Cawthon}, {Chang}, {Choi}, {De Vicente},
  {Drlica-Wagner}, {Elvin-Poole}, {Frieman}, {Gatti}, {Hartley}, {Honscheid},
  {Hoyle}, {Huff}, {Huterer}, {Jain}, {Jarvis}, {Kacprzak}, {Kirk}, {Kokron},
  {Krawiec}, {Lahav}, {Liddle}, {Peacock}, {Rau}, {Refregier}, {Rollins},
  {Rozo}, {Rykoff}, {S{\'a}nchez}, {Sevilla-Noarbe}, {Sheldon}, {Stebbins},
  {Varga}, {Vielzeuf}, {Wang}, {Wechsler}, {Yanny}, {Abbott}, {Abdalla},
  {Allam}, {Annis}, {Bechtol}, {Benoit-L{\'e}vy}, {Bertin}, {Brooks},
  {Buckley-Geer}, {Burke}, {Carnero Rosell}, {Carrasco Kind}, {Carretero},
  {Castander}, {Crocce}, {Cunha}, {D'Andrea}, {da Costa}, {DePoy}, {Desai},
  {Diehl}, {Dietrich}, {Doel}, {Fernandez}, {Flaugher}, {Fosalba},
  {Garc{\'{\i}}a-Bellido}, {Gaztanaga}, {Gerdes}, {Giannantonio}, {Goldstein},
  {Gruendl}, {Gschwend}, {Gutierrez}, {James}, {Jeltema}, {Johnson}, {Johnson},
  {Kent}, {Kuehn}, {Kuhlmann}, {Kuropatkin}, {Li}, {Lima}, {Lin}, {Maia},
  {March}, {Marshall}, {Martini}, {Melchior}, {Menanteau}, {Miquel}, {Mohr},
  {Neilsen}, {Nichol}, {Nord}, {Petravick}, {Plazas}, {Romer}, {Roodman},
  {Sako}, {Sanchez}, {Scarpine}, {Schindler}, {Schubnell}, {Smith}, {Smith},
  {Soares-Santos}, {Sobreira}, {Suchyta}, {Swanson}, {Tarle}, {Thomas},
  {Tucker}, {Vikram}, {Walker}, {Weller}, \& {Zhang}}]{DES_shear_Y1}
{Troxel}, M.~A., {MacCrann}, N., {Zuntz}, J., {et~al.} 2017, ArXiv e-prints,
  arXiv:1708.01538

\bibitem[{{Van Waerbeke} {et~al.}(2000){Van Waerbeke}, {Mellier}, {Erben},
  {Cuillandre}, {Bernardeau}, {Maoli}, {Bertin}, {McCracken}, {Le F{\`e}vre},
  {Fort}, {Dantel-Fort}, {Jain}, \& {Schneider}}]{VanWaerbeke:2000}
{Van Waerbeke}, L., {Mellier}, Y., {Erben}, T., {et~al.} 2000, \aap, 358, 30

\bibitem[{{Weinberg} {et~al.}(2013){Weinberg}, {Mortonson}, {Eisenstein},
  {Hirata}, {Riess}, \& {Rozo}}]{Weinberg:2013a}
{Weinberg}, D.~H., {Mortonson}, M.~J., {Eisenstein}, D.~J., {et~al.} 2013,
  \physrep, 530, 87

\bibitem[{{Wittman} {et~al.}(2000){Wittman}, {Tyson}, {Kirkman},
  {Dell'Antonio}, \& {Bernstein}}]{Wittman:2000a}
{Wittman}, D.~M., {Tyson}, J.~A., {Kirkman}, D., {Dell'Antonio}, I., \&
  {Bernstein}, G. 2000, \nat, 405, 143

\bibitem[{{Wong} {et~al.}(2017){Wong}, {Suyu}, {Auger}, {Bonvin}, {Courbin},
  {Fassnacht}, {Halkola}, {Rusu}, {Sluse}, {Sonnenfeld}, {Treu}, {Collett},
  {Hilbert}, {Koopmans}, {Marshall}, \& {Rumbaugh}}]{Wong:2017a}
{Wong}, K.~C., {Suyu}, S.~H., {Auger}, M.~W., {et~al.} 2017, \mnras, 465, 4895

\end{thebibliography}

\end{document}